\pdfoutput=1
\documentclass[3p,times]{elsarticle}

\usepackage{ecrc}

\volume{00}

\firstpage{1}

\journalname{Nuclear Physics A}

\runauth{}

\jid{nupha}

\usepackage{amssymb}
\usepackage[figuresright]{rotating}

\DeclareGraphicsExtensions{.pdf,.png,.jpg}
\begin{document}

\begin{frontmatter}



% \dochead{}

\title{Nuclear modification of $J/\psi$ production in Pb--Pb collisions at $\sqrt{s_{NN}}=2.76$\,TeV}


% Principle author, and co-authors - please modify
\author{Jens Wiechula for the ALICE collaboration}
\address{Physikalisches Institut der Universit\"at T\"ubingen, 72076 T\"ubingen, Germany}
\ead{Jens.Wiechula@cern.ch}

\begin{abstract}
ALICE is the dedicated heavy-ion experiment at the LHC.
Due to the unique particle identification capabilities of the central barrel detectors ($|\eta|<0.9$), $J/\psi$ can be measured in the di-electron channel
  in the very demanding environment of central Pb--Pb collisions at the LHC.
In addition $J/\psi$ are measured at forward rapidity ($2.5 < y < 4$) with a dedicated muon spectrometer.
ALICE is the only LHC experiment with an acceptance for $J/\psi$ that reaches down to $p_{\rm T} = 0$ at both, mid- and forward rapidity.
Preliminary results on the nuclear modification factor of the inclusive $J/\psi$ production at mid- and forward rapidity in Pb--Pb collisions at $\sqrt{s_{NN}}=2.76$\,TeV are presented.
\end{abstract}

\begin{keyword}
ALICE \sep quarkonia \sep nuclear modification

\end{keyword}

\end{frontmatter}

% \linenumbers

\section{Introduction}
\label{}
Heavy quarkonium states, such as the $J/\psi$, are expected to provide essential information on the properties of high-energy heavy-ion collisions where the formation of a Quark-Gluon Plasma (QGP) is expected.
The impact on the $J/\psi$ production of such a hot and dense medium formed in the early times of the collision has been extensively studied at SPS and RHIC energies.

It is expected that due to colour screening mechanisms $J/\psi$ production is suppressed in a plasma of quarks and gluons~\cite{matsui1986jpsi} and thus provides a unique probe for QGP formation.
$J/\psi$ suppression, however, is also induced by other effects that have to be taken into account, such as nuclear shadowing, cold nuclear matter effects and comover absorption scenarios.
At LHC energies charm is produced abundantly in central Pb--Pb collisions and scenarios where originally uncorrelated charm and anti-charm quarks (re)combine gain importance.
Such scenarios are described in several statistical and transport models (e.g.~\cite{Andronic:2011yq,Zhao:2011cv,Liu:2009nb}).
Measuring the $J/\psi$ production at LHC will help to disentangle between the different mechanisms.

\section{Data Analysis}
ALICE~\cite{ALICEjinst} measures the $J/\psi$ production at mid-rapidity ($|y|<0.9$) in the di-electron channel, as well as at forward rapidity ($2.5 < y < 4$) in the di-muon decay channel.

\begin{figure}[ht]
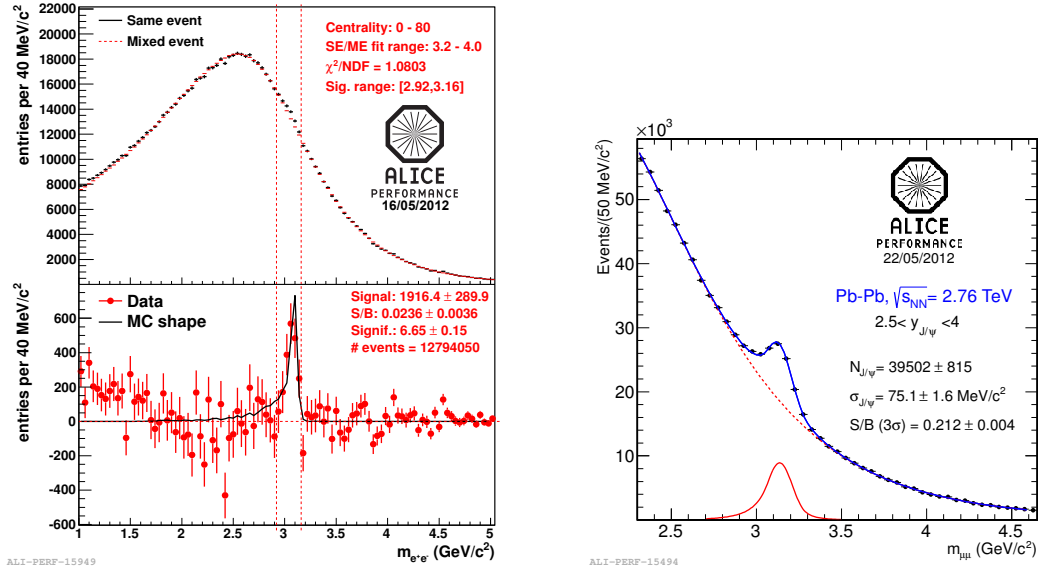

\centering
\includegraphics[width=0.4\textwidth]{{{fig1a}}}\hspace{1cm}
\includegraphics[width=0.4\textwidth]{{{fig1b}}}
\caption[]{Invariant mass spectra of di-lepton pairs. The left figure shows the di-electron channel at mid-rapidity ($|y|<0.9$), the right figure the di-muon channel at forward rapidity ($2.5 < y < 4$). For details see text.}
\label{fig1}
\end{figure}

The di-electron analysis is based on a dataset of Pb--Pb collisions at $\sqrt{s_{\rm NN}} = 2.76$\,TeV, taken with a minimum bias trigger in 2010. In total 12.8\,M (0-80\,\% most central) events were analysed corresponding to an integrated luminosity of 2.1\,${\rm \mu b^{-1}}$. Fig.\,\ref{fig1} (left) shows the invariant mass spectrum of the selected $e^+e^-$ candidates. In the top panel the opposite sign spectrum of the same event (black) is displayed together with the background distribution obtained from event mixing (red). The background distribution is scaled to match the integral of the signal distribution in an invariant mass range between 3.2 and 4\,GeV/$c^2$. The bottom panel shows the background subtracted distribution. Signal extraction is performed by integrating the signal counts in an invariant mass range of 2.92-3.16\,GeV/$c^2$. In total about 2000 $J/\psi$ are available for the analysis. This statistics allowed a signal extraction in three centrality intervals.%: 0-10\,\%, 10-40\,\% and 40-80\,\% most central events.

In case of the di-muon analysis a dataset at the same collision energy, measured in 2011, was analysed. A dedicated di-muon trigger allowed for recording 70\,${\rm \mu b^{-1}}$ of integrated luminosity, corresponding to 17.7\,M triggered events. This corresponds to a factor 20 higher statistics than previously reported~\cite{aliceFirstRaa}. Fig.\,\ref{fig1} (right) shows the invariant mass spectrum of $\mu^+\mu^-$ candidates for 0-90\,\% most central events. The signal is extracted by performing a combined fit for the background and signal contribution. The background is described by a Gaussian with a width ($\sigma$) which varies as a function of the mass value. For the signal description a modified Crystal Ball function was used which is a convolution of a Gaussian and power law functions that can fit the tails of the measured signal. In total about 40\,k $J/\psi$ have been found. This statistics allows for a differential analysis of the nuclear modification factor in either nine bins of centrality, seven bins in transverse momentum or six bins in rapidity.

The raw yield has been corrected for acceptance and efficiency. For the di-electron analysis Hijing simulations enriched with primary $J/\psi$ were used. In case of the di-muon analysis Monte Carlo (MC) $J/\psi$ were embedded into real Pb--Pb events. For the primary $J/\psi$ sample in both cases a parametrisation was used for the transverse momentum and rapidity shape, obtained from an interpolation of RHIC, Fermilab and LHC data~\cite{Bossu:2011qe}. In addition the effects of shadowing were taken into account using EKS98 calculations~\cite{eks98}. The polarisation of the simulated $J/\psi$ was assumed to be zero. First measurements of the $J/\psi$ polarisation in pp collisions at $\sqrt{s}=7$\,TeV~\cite{aliceJpsiPol} ($2.5<y<4$, $2<p_{\rm T}<8$\,GeV/$c$), show results compatible with zero.

In both cases there is only a weak dependence of the overall efficiency on the collision centrality. For the di-electron channel the total acceptance times efficiency is about 8\,\%, in case of the di-muon channel it is about 14\,\%.

% The main detectors used for the di-electron channel are the Inner Tracking System (ITS), a six layer silicon detector used for vertexing, tracking and in the trigger condition. And the Time Projection Chamber (TPC), used as the main tracking detector and for particle identification via the specific energy loss in the detector gas (d$E$/d$x$). Only tracks that are commonly reconstructed in the ITS and TPC, loosely point to the collision vertex and have a minimum $p_{\rm T}$ of $0.8$\,GeV/c are kept for the analysis. Electrons candidates are identified by requiring a three sigma compatibility with the d$E$/d$x$ expectation of electrons and rejecting tracks with a 3.5 sigma compatibility with the pion and proton expectation to reduce the background.

% In the forward direction ALICE has a dedicated muon spectrometer consisting of a tick absorber to reject hadrons, a dipole magnet, five tracking stations for the track reconstruction and a muon trigger system.

\section{Results}
The nuclear modification factor $R_{\rm AA}$ is defined as:
\begin{equation}
  R_{\rm AA} = \frac{Y^{PbPb}_{J/\psi}}{T_{\rm AA} \cdot \sigma^{pp}_{J/\psi}},
    \quad Y^{PbPb}_{J/\psi} = \frac{N^{PbPb}_{J/\psi}}{BR \cdot (A\times\epsilon) \cdot N^{events}_{MB}},
\end{equation}
where the $J/\psi$ yield, $Y^{PbPb}_{J/\psi}$, is the number of $J/\psi$ ($N^{PbPb}_{J/\psi}$) corrected for acceptance times efficiency ($A\times\epsilon$), the branching ratio ($BR$) and normalised to the number of analysed minimum bias interactions $N^{events}_{MB}$. $T_{\rm AA}$ is the nuclear overlap function, obtained from a Glauber MC, $\sigma^{pp}_{J/\psi}$ is the $J/\psi$ production cross-section in pp collisions at the same centre of mass energy per nucleon pair, which was measured by ALICE~\cite{jpsippRef}.

\begin{figure}[ht]
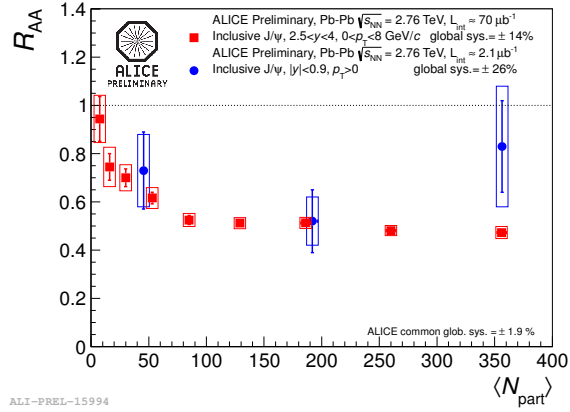

\centering
\includegraphics[width=0.45\textwidth]{{{fig2}}}
\caption[]{$R_{\rm AA}$ as a function of the average number of participant nucleons at mid- (blue) and forward (red) rapidity.}
\label{fig2}
\end{figure}

Fig.\,\ref{fig2} shows the preliminary $R_{\rm AA}$ at mid- and forward rapidity as a function of the collision centrality, expressed by the number of participant nucleons. The error bars correspond to the statistical uncertainties, the boxes to the uncorrelated systematic uncertainties of the Pb--Pb measurement. The main contribution to the systematic uncertainty in case of the di-electron channel results from the imperfections of the description of the combinatorial background. Other contributions are the uncertainty in $T_{\rm AA}$, uncertainties in the MC description of the detector and possible variations of the MC input spectrum used for the $J/\psi$. In case of the di-muon channel $T_{\rm AA}$ carries the main uncertainty, followed by the signal extraction and the trigger and tracking efficiency. Global systematic uncertainties are quoted in the figure legend: 26\,\% for the di-electron and 14\,\% for the di-muon channel, where the main contribution is due to the respective pp reference. It has to be remarked that the uncertainties in the di-electron measurement are rather large, due to the low statistics that was available for the analysis.

\begin{figure}[ht]
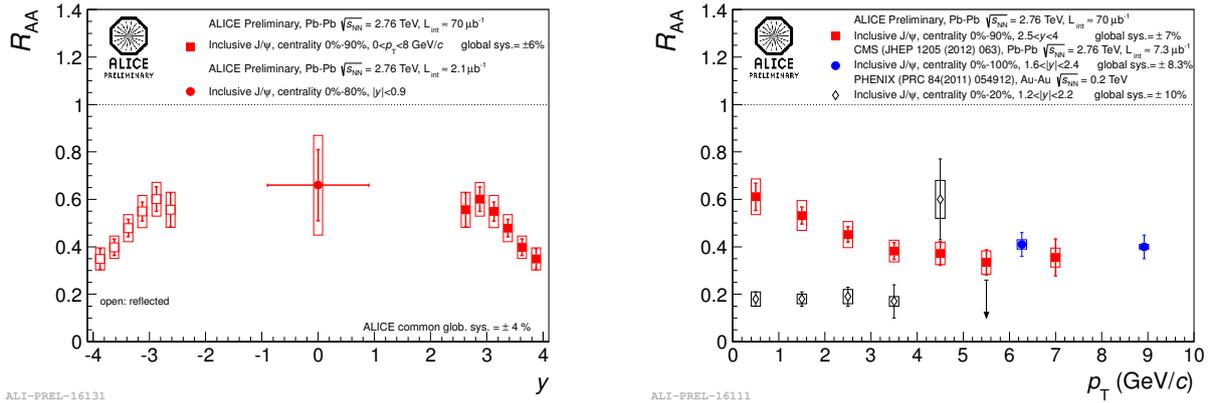

\centering
\includegraphics[width=0.45\textwidth]{{{fig3a}}}\hspace{1cm}
\includegraphics[width=0.45\textwidth]{{{fig3b}}}
\caption[]{Left panel: the nuclear modification factor as a function of rapidity measured in ALICE. Right panel: transverse momentum dependence of the nuclear modification factor measured by ALICE and CMS at the LHC in comparison to the data at RHIC.}
\label{fig3}
\end{figure}

Fig.\,\ref{fig3} (left) shows the centrality integrated $R_{\rm AA}$ as a function of rapidity. The mid-rapidity point is for 0-80\,\% most central events, the forward region corresponds to 0-90\,\% most central events. The open symbols are the forward results reflected at mid-rapidity. It can be clearly seen that $R_{\rm AA}$ decreases with rapidity.

Fig.\,\ref{fig3} (right) shows $R_{\rm AA}$ as a function of $p_{\rm T}$ as measured by ALICE (red) compared to CMS data~\cite{CMSraa} (blue) as well as results from PHENIX~\cite{phenixRaa} (black) at a lower collision energy. The $R_{\rm AA}$ is decreasing from 0.6 at low $p_{\rm T}$ to about 0.4 at higher $p_{\rm T}$. The CMS results (0-100\,\% centrality, $1.6<|y|<2.4$, $p_{\rm T}>6.5$\,GeV/$c$) are in agreement with the ALICE measurements (0-90\,\% centrality, $2.5<y<4$, $p_{\rm T}>0$) in the overlapping transverse momentum range. The lower energy results from PHENIX (0-20\,\% centrality, $1.2<|y|<2.2$) show a significantly smaller $R_{\rm AA}$.

\begin{figure}[ht]
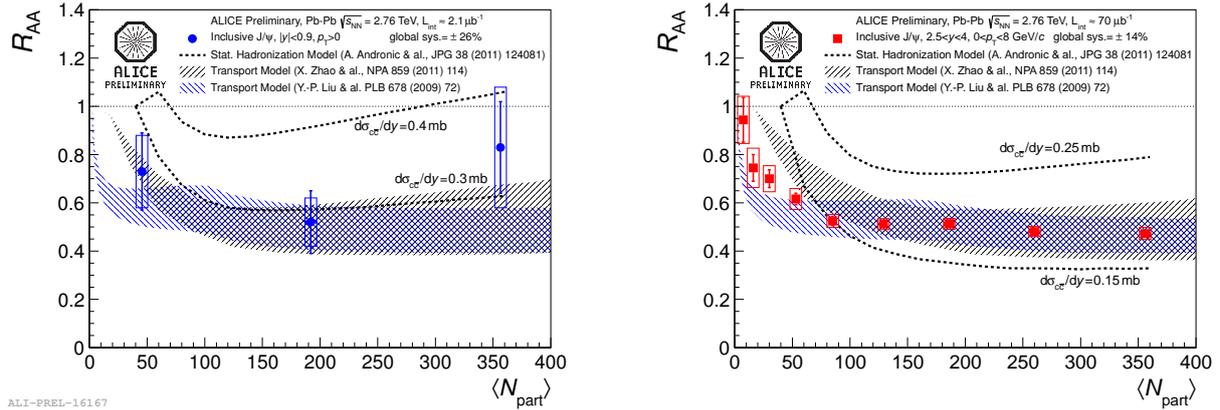

\centering
\includegraphics[width=0.45\textwidth]{{{fig4a}}}\hspace{1cm}
\includegraphics[width=0.45\textwidth]{{{fig4b}}}
\caption[]{$R_{\rm AA}$ as a function of the average number of participants at mid- (left) and forward (right) rapidity in comparison with model calculations that take into account (re)combination of $c$ and $\bar{c}$ quarks.}
\label{fig4}
\end{figure}

Fig.\,\ref{fig4} compares the ALICE results at mid- (left) and forward (right) rapidity with results from a statistical hadronization model~\cite{Andronic:2011yq}, as well as two different transport models~\cite{Zhao:2011cv,Liu:2009nb}. All models take into account (re)combination of $c\bar{c}$-pairs. Within the uncertainties the models describe the data for $N_{\rm part}$ larger than 50.

\section{Conclusions}
ALICE has measured the nuclear modification factor ($R_{\rm AA}$) of inclusive $J/\psi$ production in Pb--Pb collisions at $\sqrt{s_{\rm NN}}=2.76$\,TeV at mid- and forward rapidity. In comparison with results at RHIC energies, the $R_{\rm AA}$ at LHC energies is significantly larger. The CMS results, measured at slightly lower rapidity, are in agreement with the ALICE data in the overlapping transverse momentum range ($p_{\rm T} > 6$\,GeV/$c$). It was shown that $R_{\rm AA}$ decreases with increasing $p_{\rm T}$ and towards larger rapidity. A comparison with several models that take into account a (re)combination of $c\bar{c}$-pairs are in agreement with the ALICE results.

%% References


% \bibliographystyle{elsarticle-num}
% \bibliography{Wiechula}
 % do not change






% \end{thebibliography}

\end{document}